\documentclass{mem}
\usepackage{natbib}\usepackage{txfonts}\usepackage{balance}
\usepackage{graphicx}
\usepackage[a4paper]{hyperref}
\idline{75}{282}

\usepackage[]{wasysym}
\usepackage{subfigure}
\usepackage{tabularx}
\usepackage{amssymb}
\usepackage{epsfig}
\usepackage{wrapfig}

\begin{document}
\def\teff{$T\rm_{eff }$}
\def\kms{$\mathrm {km s}^{-1}$}

\title{
A Quasar Wind Model
}

\author{
A. \,Ruff\inst{1}
          }

\institute{
University of Melbourne
Parkville 3052, 
Victoria,
Australia \hspace{25mm} \\
\email{aruff@unimelb.edu.au}
}

\authorrunning{Ruff }

\titlerunning{A Quasar Wind Model}

\abstract{
A quasar wind model is proposed to describe the spatial and velocity structure of the broad line region. 
This model requires detailed photoionization and magnetohydrodynamic simulation, as the broad line region 
it too small for direct spatial resolution. 
The emission lines are Doppler broadened, since the gas is moving at high velocity. 
The high velocity is attained by the gas from a combination of radiative and magnetic driving forces. 
Once this model is complete, the model predictions will be tested against recent microlensing data in conjunction 
with diverse existing observations.  

}

\maketitle{}

\section{Introduction}
Broad emission lines are a prominent spectral features of quasars. 
The physical state and geometry of this region is 
poorly understood since the broad line region is not directly resolvable. 

The broadening of these emission lines is so extreme, that speeds of up to 0.1$c$ are observed. 
Temperature and number density limits can be inferred from observations of the relative strengths of broad emission lines. 
The temperature inferred from observations is too low for the broadening process to be thermal, therefore the gas 
producing these emission lines is moving at high speed relative to the quasar rest frame. 

This model proposes that the gas is accelerated outward by photon and magnetohydrodynamic forces. 
Direct spatial resolution of this region is not possible. 
In order to determine the geometry, origin and nature of this gas, simulation is required.

\section{Photoionization}
The model aims to simulate the broad emission line spectrum of a quasar using photoionization. 
The photoionization code, {\em Cloudy} (written by Gary Ferland and collaborators) will be used to calculate both 
the radiative forces accelerating the gas and a spectrum from a given set of initial conditions. 
Free model parameters can be constrained by varying the initial input conditions. 

Asymmetrical profiles of broad emission lines are strong evidence for an outflowing wind \citep{elvis}. 
In the case of an outflowing gas, the line optical depths are modified by
\begin{equation}
\tau_l = \kappa_l \rho_{gas} v_{thermal} |a_{gas}|^{-1}
\end{equation}
as described by \citet{castor}, where $\tau_l$ is the line optical depth, $\kappa_l$ is the mono-chromatic line absorption co-efficient, 
$\rho_{gas}$ is the broad line gas density, $v_{thermal}$ the thermal velocity of the gas and $a_{gas}$ the gas acceleration. 

\section{Accelerating the Gas}
\begin{figure*}[t!]
\resizebox{\hsize}{!}{\includegraphics[width= 0.49\textwidth]{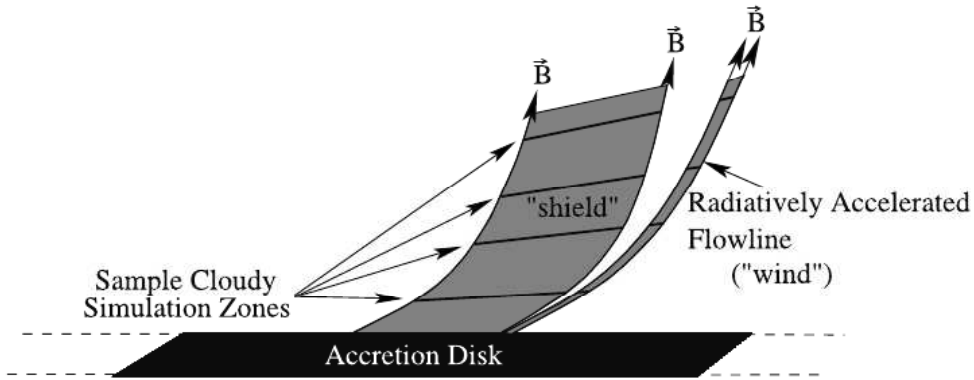}}
\caption{\footnotesize
A schematic showing the proposed structure of the wind. \citet{everett}
}
\label{fig}
\end{figure*}

In this model, the gas that produces the broad emission lines originates from the accretion disk. 
The gas is then accelerated outward magnetohydrodynamically and radiatively by both continuum and line driving. 
X-ray shielding gas is included in this model, which prevents the wind from becoming over-ionized, as shown in Figure \ref{fig}. 

\section{Tests of this model}

Broad line flux ratios are highly indicative of physical parameters such as number density, ionization parameter and metallicity. 
Observations of broad line flux ratios can be used to constrain free parameters and test the validity of this model. 
{\em Cloudy} simulations also predict the flux of each broad emission line as a function of radius. 
These predictions can then be compared to reverberation mapping results as a test of the model predictions. 
It is possible to extend this model to account for gas clumping, rather than a smooth outflow as described. 
It is expected that a broad line region consisting of clumped gas would give significantly different results, 
since the velocity profile and ionization structure would differ. 


\bibliographystyle{aa}

\end{document}